\documentclass[12pt]{iopart}
\usepackage{iopams}
\usepackage{cite}
\usepackage{braket}
\usepackage{graphicx}
\usepackage{color}
\usepackage{epstopdf}
\usepackage{citesort}
\pdfoutput=1
\newcommand{\bS}{\textbf{S}}
\newcommand{\vk}{\textbf{k}}

\newcommand{\eqref}[1]{(\ref{#1})}
\usepackage{soul}

\begin{document}

\title[Topological order in the Haldane model with spin-spin on-site interactions]{Topological order in the Haldane model with spin-spin on-site interactions}

\author{A. Rubio-Garc\'ia$^1$ and J. J. Garc\'ia-Ripoll$^2$}

\address{$^1$ Instituto de Estructura de la Materia IEM-CSIC, Calle Serrano 123, Madrid 28006, Spain}
\ead{alvaro.rubiog@iem.cfmac.csic.es}
\address{$^2$ Instituto de F\'isica Fundamental IFF-CSIC, Calle Serrano 113b, Madrid 28006, Spain}

\begin{abstract}
Ultracold atom experiments allow the study of topological insulators, such as the noninteracting Haldane model. In this work we study a generalization of the Haldane model with spin-spin on-site interactions that can be implemented on such experiments. We focus on measuring the winding number, a topological invariant, of the ground state, which we compute using a mean-field calculation that effectively captures long range correlations and a matrix product state computation in a lattice with 64 sites. Our main result is that we show how the topological phases present in the noninteracting model survive until the interactions are comparable to the kinetic energy. We also demonstrate the accuracy of our mean-field approach in efficiently capturing long-range correlations. Based on state-of-the-art ultracold atom experiments, we propose an implementation of our model that can give information about the topological phases.
\end{abstract}

\noindent{\it Keywords\/}: topological phases, topological insulators, Hubbard interactions, ultracold atoms

\section{Introduction}\label{sec:introduction}

Ultracold atoms trapped in optical lattices allow the simulation of rich physical models that cannot be easily recreated in solid state materials\  \cite{bloch05,bloch08}. Great experimental breakthroughs include the simulation of the Bose-Hubbard\ \cite{greiner02} and Fermi-Hubbard models\ \cite{jordens08,schneider08}, the implementation of fermionic and bosonic lattice models with artificial gauge fields, such as the Hofstadter model\ \cite{aidelsburger13,hirokazu13,kennedy15} or topological Floquet band models\ \cite{flaschner16}, and, of particular relevance for this work, the realization of the Haldane model\ \cite{jotzu14}. The Haldane model\ \cite{haldane88} is an example of a 2D topological insulator (TI), a linear model with a nontrivial topological invariant which, while being insulating in their bulk, supports protected currents along its edges\ \cite{hasan10}. An open problem in the study of topological matter, and the focus of this work, is the interplay between interactions and the topological characterization of the system. An ambitious goal would be to understand why and how ultrastrong interactions can convert a topological insulator into a topologically ordered quantum state of matter, which features degenerate, locally indistinguishable ground states and topologically nontrivial excitations. However, before achieving this goal, it would be very interesting to focus on the first steps of this transition, studying the persistence of topologically nontrivial phases in the presence of interactions.

Some work has already been done in the field of interacting topological insulators, such as the Haldane-Hubbard model in the Mott insulator limit\ \cite{arun16,hickey15,hickey16,gu15,imriska16,wu15,wu16}, the Hubbard model with nearest neighbor interactions\ \cite{alba16}, the synthetic Creutz-Hubbard model\ \cite{junemann16} or a general TI in the context of the Fractional Quantum Hall Effect\ \cite{wu12}. Our novel contribution in this work is to study the topological phases of the Haldane model with spin $1/2$ fermions for a broad range of on-site interactions. This model, which represents a straightforward extension of recent experiments with noninteracting fermionic atoms\ \cite{jotzu14}, is studied using two methods: (i) a mean-field approach in momentum space that effectively takes into account long range correlations and (ii) a matrix product state (MPS) ansatz in a 2D lattice\ \cite{white92,schollwock11} with 64 sites. We have found that the topological phases which appear in the noninteracting model extend to values of the interaction which are comparable to the kinetic energy of the atoms in the lattice. For stronger interactions, the ground state becomes a Mott insulator, where, in accordance to earlier work, a nontrivial spin order may appear\ \cite{arun16,hickey15,hickey16,gu15,imriska16,wu15,wu16}. On account of our numerical results, we propose an ultracold atom experiment that can directly simulate our model, based on previous experiments with TIs\ \cite{jotzu14,aidelsburger13,hirokazu13} and the measurement of topological properties\ \cite{tarruel12}.

The article is structured as follows. In section \ref{sec:the_model} we describe a generalization of the Haldane model with spin-spin on-site interactions, how topology arises in the noninteracting limit and how it can be detected by measuring the winding number. In section \ref{sec:mean_field} we begin our study of interactions, introducing a mean-field variational wavefunction that exactly reproduces the noninteracting ground state. This function is a product state in momentum state and may capture the long-range correlations needed to describe a topological phase with interactions. Using a global optimization procedure, we estimate the ground state energy and wavefunction within this ansatz. We find evidence of a topological phase for nonzero interactions, as well as a cross over into a Mott insulator and double occupancy regions for strongly repulsive and strongly attractive interactions, respectively. In section\ \ref{sec:MPS} we study the same problem, using the MPS ansatz spanned over a 2D honeycomb lattice. Heavy simulations with up to $64$ lattice sites confirm the predictions of the mean-field ansatz, including the topological phase transition and cross-overs, and show the accuracy of this simple wavefunction when estimating the ground state energy. Section \ref{sec:experimental_setup} discusses the modifications needed to implement our model using state-of-the-art experiments\ \cite{jotzu14}, including state preparation and detection of topological and trivial phases. We close this work with a brief summary and discussion in section \ref{sec:discussion}.

\section{The model}\label{sec:the_model}

\begin{figure}[tbp]
	\centering
	\includegraphics[scale=1]{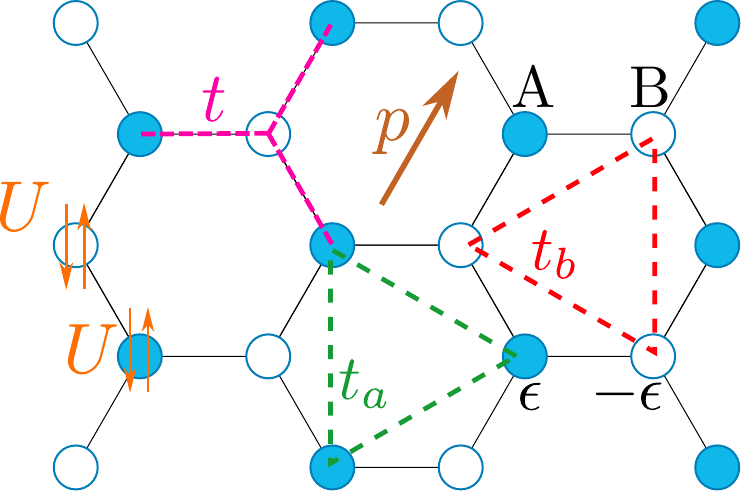}
	\caption{Scheme of the honeycomb lattice with sublattices A and B in colors blue and white, respectively. Parameters\ $t$,\ $t_a$ and\ $t_b$ stand for the hopping amplitudes and\ $p$ for the external phase.\ $\epsilon$ represents the energy imbalance between sublattices and\ $U$ is the on-site interaction energy between particles with opposite spin.} \label{fig:lattice}
\end{figure}

The Haldane model is a Hamiltonian that describes the motion of particles in a honeycomb lattice with nearest- and next-to-nearest-neighbor hopping amplitudes. The topological nature of the model arises from the existence of complex hopping amplitudes, which in our work we place in the nearest-neighbor hopping. The honeycomb lattice is composed of two triangular lattices, which we denote $A$ and $B$. Each site will be able to host up to two fermionic particles in state  $s\in\{\uparrow,\downarrow\}$, which we denote with creation operators $a_{i,s}^\dagger$ and $b_{i,s}^\dagger$ for the two sites of the $i$-th unit cell. The full model describing this system reads
\begin{equation}
  H = H_{0\uparrow} + H_{0\downarrow} + U\sum_i \left(a^\dagger_{i\uparrow}a^\dagger_{i\downarrow}a_{i\downarrow}a_{i\uparrow}+
    b^\dagger_{i\uparrow}b^\dagger_{i\downarrow}b_{i\downarrow}b_{i\uparrow}\right),
  \label{eq:Haldane}
\end{equation}
where $H_{0,s}$ is the kinetic energy of each fermionic species, and $U$ is the on-site interaction between distinguishable fermions.

The kinetic part has the usual form
\begin{equation}
\fl H_{0s} = 
-\sum_{\langle i,j \rangle} \left( t \rme^{i\phi_{i j}} a_{i,s}^\dagger b_{j,s} +h.c. \right) 
-\sum_{\langle\langle i,j \rangle\rangle} \left( t_a a_{i,s}^\dagger a_{j,s}+t_b b_{i,s}^\dagger b_{j,s} \right)  
-\sum_{i} \epsilon\left( a_{i,s}^\dagger a_{i,s}-b_{i,s}^\dagger b_{i,s}\right).
\label{eq:hopping_hamiltonian}
\end{equation}
This noninteracting model contains a first-neighbor hopping, $t$, which is affected by a phase that grows linearly along the lattice direction $\textbf{p}$, $\phi_{ij} = \textbf{p}\left(\textbf{x}_i - \textbf{x}_j\right)$. There are also two next-to-nearest neighbor hopping amplitudes $t_a$ and $t_b$, on the respective sublattices (see figure\ \ref{fig:lattice}); and an on-site energy imbalance between lattices $\epsilon$. In this work we have set: $t=1$,  $t_a =  -t_b = 0.1$ and $\epsilon=0$.

The hopping Hamiltonian for each spin population \eqref{eq:hopping_hamiltonian} is a two band model that can be written as an effective magnetic field acting in momentum space. Introducing the Fourier transformed operators\ $a_{\vk,s}$ and\ $b_{\vk,s}$, and the pseudospin operators\ $\hat\sigma^x=a^\dagger b + b^\dagger a$,\ $\hat\sigma^z= a^\dagger a - b^\dagger b$ and\ $\hat\sigma^y = -i[\sigma^z,\sigma^x]$, we find
\begin{equation}
H_{0s} = \sum_\vk \textbf{B}\left(\vk\right)\hat{\bsigma}_{\vk,s}, \;s\in\{\uparrow,\downarrow\}.
\end{equation}
The field\ $\textbf{B}(\vk)$ determines a preferred orientation of the expectation value for the single-particle operators\ $\bsigma^i$. The ground\ state $\Psi_0$ of\ $H_{0s}$ is a half-filled state, for which the pseudospin field,\ $\bS_s\left(\vk\right)$, which is proportional to the expected value of the pseudospin operators,\ $\bsigma_{\vk,s}$, satisfies
\begin{equation}
\bS_s\left(\vk\right) := \frac{1}{2} 
\braket{\Psi_{0} |\bsigma_{\vk,s}
|\Psi_{0}} \propto - \textbf{B}(\vk).
\end{equation}
The direction and amplitude of the field $\textbf{B}\left(\vk\right)$ and of the pseudospin $\bS_s$ are given by
\begin{eqnarray}
\textbf{B}(\vk) =
\left( 
\mathrm{Re}\left[f\left(\vk\right) \right], 
\mathrm{Im}\left[f\left(\vk\right) \right],
\epsilon + \left(t_a - t_b\right)g\left(\vk\right)
\right)
\\
f\left(\vk\right) =
\sum_{i=1,2,3} \rme^{-i\left( \textbf{p} - \vk\right)\textbf{v}_i} \\
g\left(\vk\right) =
\sum_{i=1,2,3} \cos \left( \vk \textbf{w}_i \right),
\end{eqnarray}
where $\textbf{v}_i$ and $\textbf{w}_i$ are the vectors that connect a site with its nearest- and next-to-nearest-neighbors, respectively. The half-filled ground state without interactions is therefore completely determined by the direction of the pseudospin fields\ $\bS_s\left(\vk\right)$ of both spin populations over the Brillouin zone.

An interesting property of the hopping Hamiltonian (\ref{eq:hopping_hamiltonian}) is that different configurations of the vector field $\bS_s$ can be regarded as topologically distinct. More precisely, $\bS_s(\vk)$ may be regarded as a mapping from the Brillouin zone onto the sphere, $|\bS_s|=1$. These mappings may be classified by the  number of times the pseudospin $\bS_s$ wraps around the sphere, what is called the winding number\ \cite{deLisle14}.
\begin{equation}
\nu_s = 
\frac{1}{4\pi} 
\int \bS_s\left(\vk\right) 
\left[
\partial_{k_x}\bS_s\left(\vk\right) \times \partial_{k_y}\bS_s\left(\vk\right)
\right]
\rmd^2\vk.
\label{eq:winding}
\end{equation}
Whenever the band wraps one or more times around the sphere, the model is said to be topologically non-trivial. This results in a nonzero Chern number and the existence of topologically protected edge currents when the model is embedded in a finite domain with boundaries or holes.

A very important open question is what happens to the topological insulator model\ \eqref{eq:Haldane} when there are active interactions between particles, $U\neq 0$. In such situations, the single-particle picture is no longer valid, but the system may still exhibit topological properties, such as a nontrivial pseudospin pattern $\bS_s(\vk)$, which gives rise to different phases, or the existence of edge states, or, in the limit of infinitely strong interactions, perhaps the appearance of true topological order.

In this work we will analyze\ \eqref{eq:Haldane} using different variational methods ---mean-field theory and matrix-product states---, with the goal of classifying the approximate ground states of the model. Our main tool in doing this will be the use of the winding number, $\nu$, because it is a topological property that can be experimentally and numerically determined, without access to the wavefunctions or parallel transport.

\section{Mean-field}\label{sec:mean_field}

We have seen above that the ground state at half filling and without interactions is uniquely determined by the pseudospin field $\bS_s(\vk)$ over the Brillouin zone, with one particle per unit cell in momentum space. It makes sense, therefore, to approximate the ground state at nonzero interactions by a product state in momentum space
\begin{equation}
\ket{\Psi_0\left[\bS_\downarrow,\bS_\uparrow\right]} =
\prod_{\vk\in BZ} c_{\vk,\downarrow}^\dagger\left[ \bS_\downarrow\right] c_{\vk,\uparrow}^\dagger\left[ \bS_\uparrow\right]\ket{0}.
\end{equation}
This ground state is constructed by placing one particle per spin in each of the momentum states,\ $c_{\vk,s}^\dagger[\bS_s]= \alpha_s[\bS_s]a_{\vk,s}^\dagger + \beta_s[\bS_s]b_{\vk,s}^\dagger$. These modes are characterized by the fact that\ $\frac{1}{2} \braket{0|c_{\vk,s}\hat{\bsigma}_{\vk,s} c_{\vk,s}^\dagger|0}=\bS_s(\vk)$, where $\bS_s(\vk)$ is a vector of norm\ $1/2$ and our variational parameter. All the information about the ground state is thus given by two independent pseudospin fields $\bS_\uparrow$ and $\bS_\downarrow$, with uniform norm. The variational energy of this wave function then reads as the functional
\begin{equation}
E\left[\bS_\downarrow,\bS_\uparrow\right] = 
- 2\sum_\vk \textbf{B}_\vk \bS_{\vk\uparrow}
- 2\sum_\vk \textbf{B}_\vk \bS_{\vk\downarrow}
+ 2\frac{U}{N} \sum_{\vk\vk'} S^z_{\vk\uparrow} S^z_{\vk'\downarrow}
+ \frac{U}{2},
\end{equation}
where $N$ is the total number of unit cells in our lattice. We remark that, in the absence of interactions $U=0$ this mean-field approach is \textit{exact}.

The form of the variational energy suggests that, when the interactions are weak enough compared to the kinetic energy terms, $|U|\ll |B|$, the topological phase will remain unaffected, and the ground state will adopt a spin order determined by the effective magnetic field $\bS_{\vk,\downarrow}=\bS_{\vk,\uparrow}\propto - \textbf{B}_\vk$.  For strong repulsive interactions $U\gg|B|$, the mean field  model suggests that the vector field develops an antiferromagnetic order, where all spin components point along the $Z$ direction, $S^z_\downarrow = -S^z_\uparrow$. If we inspect the expression for the pseudospin
\begin{equation}
\fl S_{\vk,s}^z =
\frac{1}{2}\left(\braket{a_{\vk,s}^\dagger a_{\vk,s}} - \braket{b_{\vk,s}^\dagger b_{\vk,s}}\right) \sim
\sum_i \braket{a_{i,s}^\dagger a_{i,s}} - \braket{b_{i,s}^\dagger b_{i,s}},\; |U|\gg t,
\end{equation}
we realize that this implies a phase separation between spin components, whereby different spin orientations sit on different sublattices, i.e particles $\uparrow$ and $\downarrow$ are placed on sublattices $A$ and $B$, respectively, or vice versa. Moreover, because the transverse components are zero $S^{x,y}\simeq 0$, the winding number vanishes in this phase. Similarly, for strong interactions, $U\ll -|B|$, we expect a ferromagnetic order to develop, which amounts to having bunching of particles on the same lattice sites.
 
\begin{figure}[tbp]
	\includegraphics[width=\textwidth]{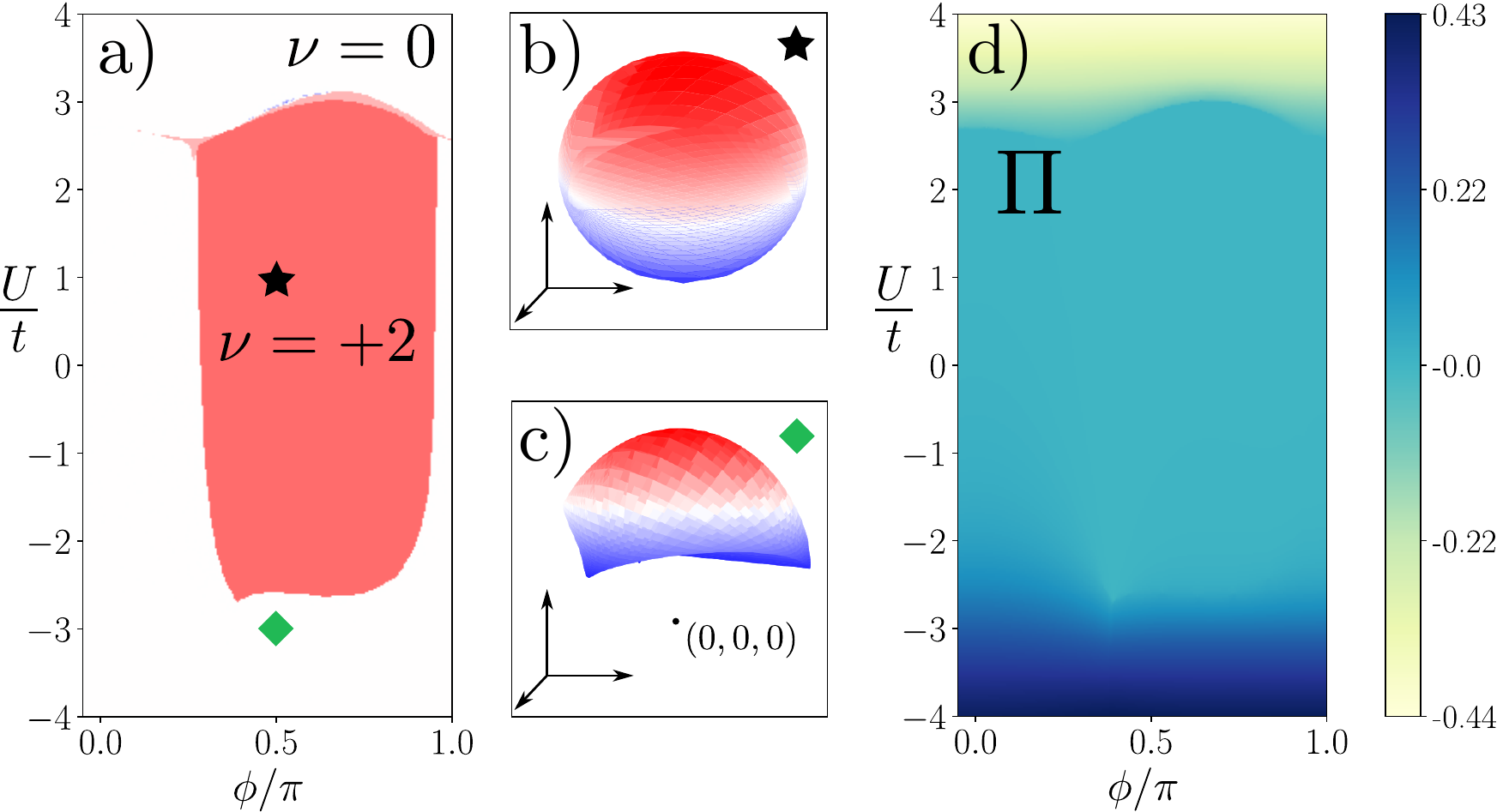}
	\caption{Mean-field results of a lattice with $20\times 20$ unit cells. $(a)$ Topological phases in the mean-field simulation signaled by the winding number. $(b)$ and $(c)$ Map of the pseudospin field onto the Bloch sphere at points in phase space shown in $(a)$. $(b)$ represents a pseudospin configuration that wraps the whole sphere, therefore having\ $|\nu_b| = 1$; while the pseudospin configuration in $(c)$ does not wrap around a whole sphere, thus\ $\nu_c=0$. $(d)$ Order parameter $\Pi$ \eqref{eq:phase_separation}.} \label{fig:mean_field}
\end{figure}

We have optimized numerically the variational energy for lattices with up to 400 sites, using a linear search algorithm that starts with a random distribution of vector fields. For each of the variational ground states, we have computed both a discrete approximation to the winding number\ \eqref{eq:winding}, as well as an order parameter that detects phase separation and bunching. In the figure\ \ref{fig:mean_field}a we plot the winding number of the mean-field ground state of a lattice with $20 \times 20$ unit cells. We see that the topological phases survive until the strength of the interactions, $|U|$, is around three times bigger than the kinetic energy, $t$. In \ \ref{fig:mean_field}b we plot the order parameter coming from the expression of the variational mean-field energy,
\begin{equation}
  \Pi := \frac{4}{N^2}\sum_{\vk\vk'} S^z_{\vk,\uparrow} S^z_{\vk',\downarrow}.
  \label{eq:phase_separation}
\end{equation}
In regions which correspond to a topological phase,\ $\Pi$ is strictly zero, but it breaks into two distinct phases when the topological order disappears. It then becomes negative in the case\ $U > 0$, when the interactions are repulsive, which means that particles of opposite spin sit on different sublattices, therefore signaling a charge density wave. When interactions are attractive,\ $U < 0$, the expectation value\ $\Pi$ becomes positive, meaning that all particles sit on the same sublattice.

\section{Matrix product state simulations}
\label{sec:MPS}

As a check of the validity of our mean-field approach we have used an MPS ansatz to compute the ground state of the interacting model and the winding number. MPS is a powerful method to compute the ground state of quantum lattice models that relies on the low scaling of the entanglement entropy in ground states in 1D and 2D. It describes a quantum state as a product of tensors on every lattice site
\begin{equation}
\ket{\Psi} = 
\sum_{i_0\ i_1\ \cdots}
A^{i_0}\left[0\right] A^{i_1}\left[1\right] \cdots
\ket{i_0 i_1 \cdots}
\end{equation}
where the dimension of the $i_j$ indices are the physical dimensions of the system at the $j$-th lattice site, and $A^{i_j}[j]$ are the tensors corresponding to the same site, whose dimensions depend on the bipartite entanglement of the state around $j$. This ansatz is an exact representation of every quantum state for sufficiently large tensors. However, for practical applications one usually sets a maximum size,  called bond dimension, $\chi$, for every $A^{i_j}[j]$, discarding configurations that contribute the least to total the ground state.

\begin{figure}[tbp]
	\centering
	\includegraphics[scale=0.6]{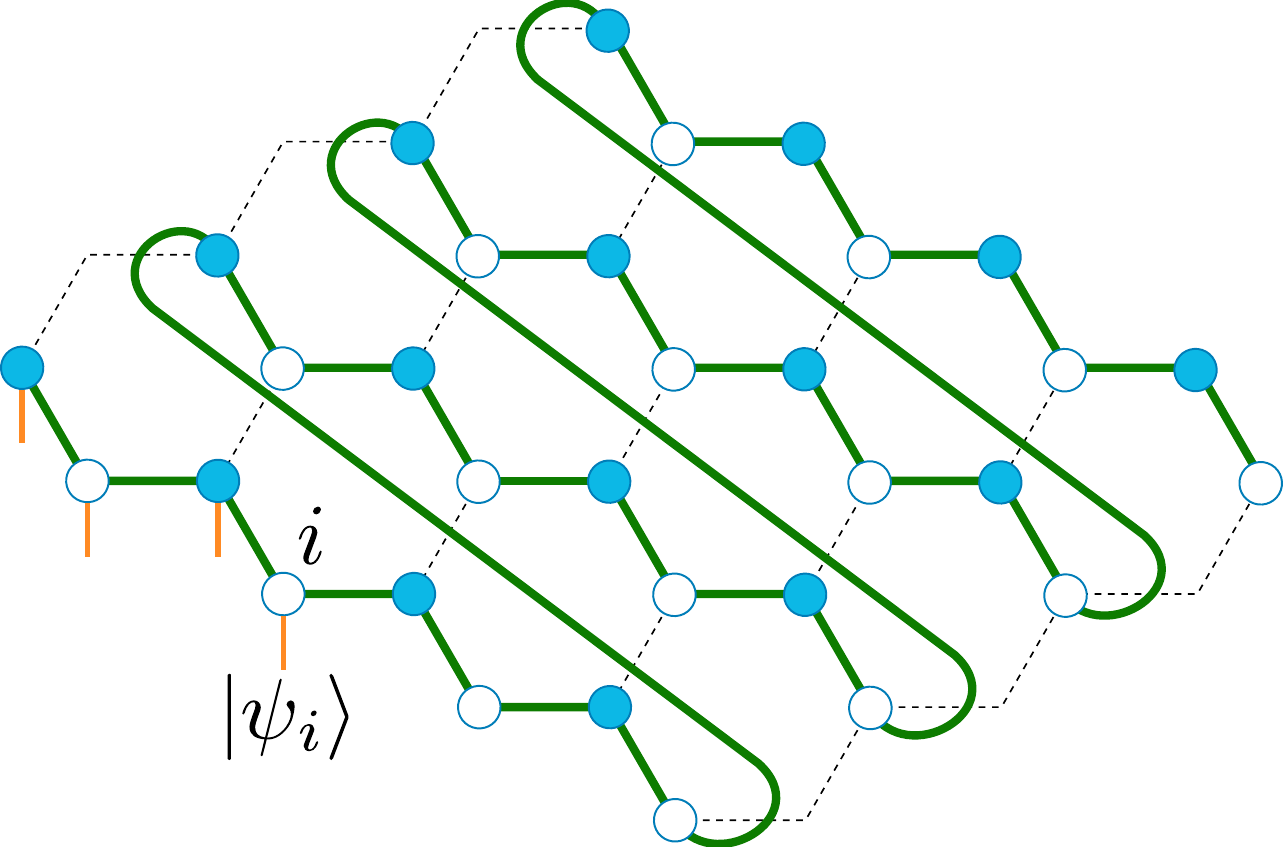}
	\caption{Scheme of our ``snake'' MPS spanning over a $4 \times 4$ unit cell honeycomb lattice with spin. The order in which the physical indices are covered is indicated by the thick green line. The MPS goes first through the lattice with spin up and then through the lattice with spin down.} \label{fig:mps_lattice}
\end{figure}

In our calculations we have constructed an MPS by mapping a lattice with $N\times N$ unit cells\ \cite{white92,schollwock11} to two consecutive 1D systems in a ``snake'' configuration, with a total of $2\times 2\times N^2$ sites, one for each spin component [cf. figure\ \ref{fig:mps_lattice}]. We have focused our efforts in diagonalizing a $4\times 4$ lattice using a maximal bond dimension of $\chi=120$. Lattices smaller than $4\times 4$ unit cells did not output correct results for the winding number, already in the noninteracting analytical solution. Bigger lattices demanded a too large bond dimension that exceeded our computational resources.

As hinted in our mean-field computation (figure\ \ref{fig:mean_field}), the nontrivial topological phases extend until the interactions are comparable to the kinetic energy,\ $|U| \sim 3t$. For stronger interactions, the term\ $U\sum_{i\in A,B} n_{i\uparrow} n_{i\downarrow}$ becomes dominant and nonlocal expectation values ($\braket{c_i^\dagger c_j},\ i\ne j$) approach zero, where different ferro- or antiferromagnetic orders appear, which correspond to different distributions of the spin components in the two sublattices. We identified the double occupancy as an appropriate order parameter to measure these distribution effects
\begin{equation}\label{eq:D}
D = 
\frac{\sum_i \braket{n_{i,\uparrow} n_{i,\downarrow}}}
{\sum_{i,s} \braket{n_{i,s}}}.
\end{equation}

\begin{figure}[tbp]
	\includegraphics[width=\textwidth]{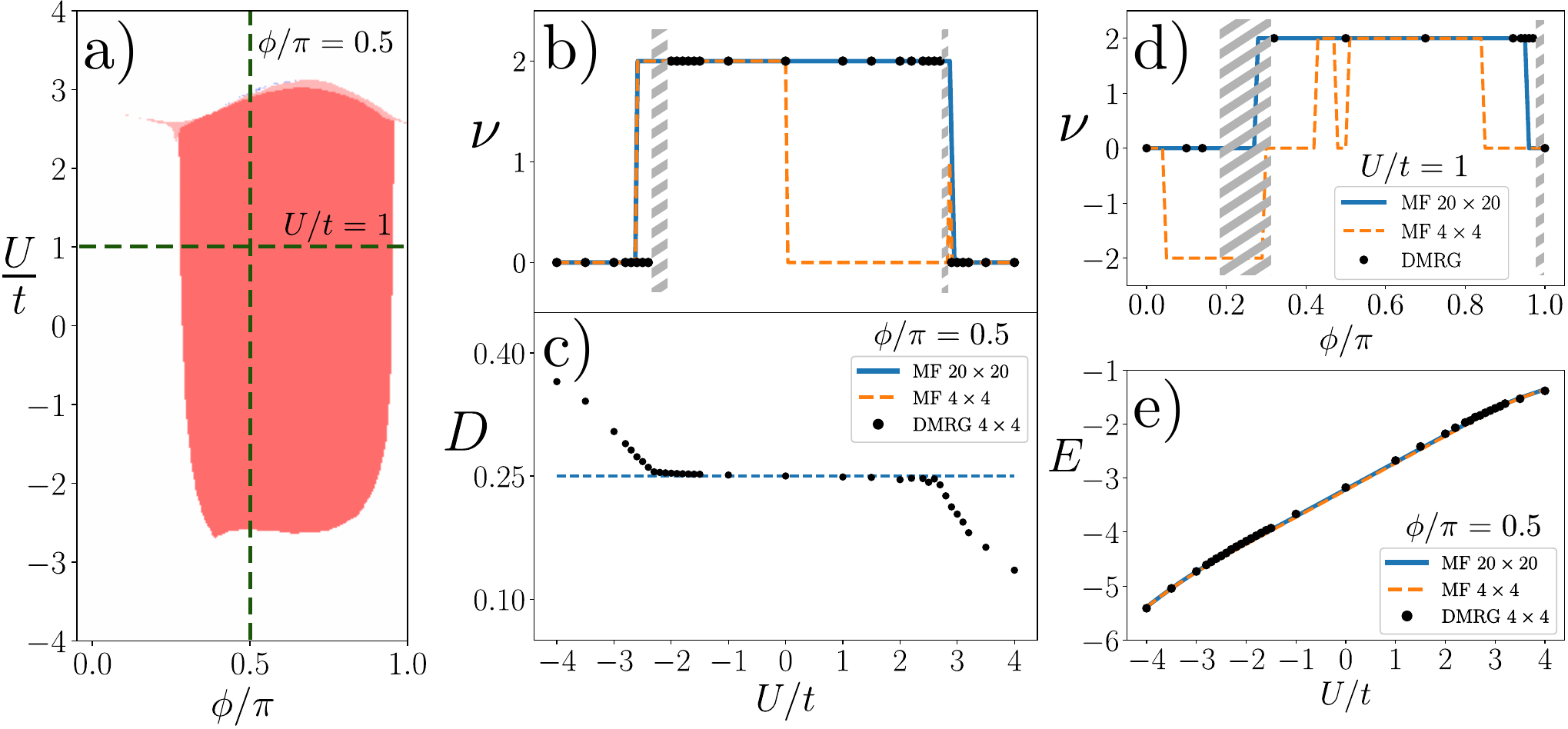}
	\caption{Results from the MPS computation with a\ $4\times 4$ unit cell lattice at\ $\chi=120$.\ $(a)$ Lines in phase space at which we have computed the winding number of the ground state.\  $(b)$ Winding number at\ $\phi/\pi=0.5$. The lines represent mean-field results with different lattice sizes while points represent our MPS computation. The shaded areas are regions where the winding computation does not converge.\ $(c)$ Double occupancy of the points at\ $\phi/\pi=0.5$.\ $(d)$ Winding number at the line\ $U/t=1$. The double occupancy at this line is constant,\ $D=0.25$. The symbols are the same as in figure\  $(b)$.\ $(e)$ Energy of the mean field computation compared to the MPS energies.} \label{fig:mps_results} 
\end{figure}

In figure\ \ref{fig:mps_results} we show the main results of our MPS simulations. Due to the big computational effort needed to get the MPS ground state at every point we have chosen to compute them only along some relevant lines in phase space, denoted by dashed lines in figure\ \ref{fig:mps_results}a. The winding numbers along these lines are shown in figures\ \ref{fig:mps_results}b and \ \ref{fig:mps_results}d. These winding numbers have been computed by extracting the expectation values\ $\braket{c_i^\dagger c_j}$ from the MPS and then Fourier transforming them into bigger lattices of\ $8\times 8$,\ $12\times 12$ or\ $16\times 16$ unit cells in momentum space. This Fourier interpolation of the pseudospin field, allows us to make a more accurate determination of the winding number in momentum space. We only keep simulation results that have converged under this interpolation procedure. The selected values compare favorably with the mean-field predictions for large lattices\ $20\times 20$, away from the phase transition.

Near the topological phase transitions, the winding number fails to converge in the MPS simulations [cf. gray shaded areas in figure\ \ref{fig:mps_results}b,d]. However, this is not surprising. As a mean-field shows, it is difficult to determine accurately the winding number with just\ $4\times 4$ pseudospins [cf. yellow line in figure\ \ref{fig:mps_results}b,d]. Around these regions, the gap closes, and it is therefore very difficult for an algorithm to determine whether the pseudospin field maps into a closed or an open surface, respectively figures\ \ref{fig:mean_field}b and c.

In figure\ \ref{fig:mps_results}c we plot the double occupancy of the ground state for the MPS computation and the mean-field at the line\ $\phi/\pi=0.5$. The expectation value of this observable should approach\ $1/4$ when the interactions have no effect on the ground state and the two topological insulators are uncoupled\ (\ref{eq:hopping_hamiltonian}). However,\ $D$ should become\ $1/2$ in the strongly attractive phases and\ $0$ in the strongly repulsive phases, where particles with opposite spin are either grouped or separated, respectively. We observe a good agreement between both simulations, specially at the plateau\ $D=1/4$, where our results show a non-trivial topological phase.

Finally, figure\ \ref{fig:mps_results}e shows the ground state energy of the mean-field simulation, together with that of the MPS computation. The mean-field energy is in a remarkably good agreement with MPS for all parameters, and it is even slightly below for\ $|U|\leq |t|$. While surprising, this can be justified by the fact that our momentum-space mean-field ansatz is nonlocal, and contains long range correlations that can be challenging to capture with the MPS, one-dimensional ansatz.

\section{Experimental setup}\label{sec:experimental_setup}

Ultracold atoms trapped in optical lattices provide an excellent tool to study the topological properties of the Haldane model, in any of its flavors. This has been demonstrated by the recent experimental work in\ \cite{jotzu14}. A crucial step to realize topological order in the lattice, is to implement an effective gauge field on the phases of the hopping amplitudes when a particle moves through the lattice. There are two essential ways in which this is done. The first one is through laser assisted hopping\  \cite{hirokazu13,aidelsburger13}, in which the lattice is divided into two sublattices which host atoms in different hyperfine states. A laser beam then transfers atoms between sublattices, while imparting a complex phase. A second way to make complex hoppings is to periodically shake the lattice\ \cite{goldman16,jotzu14}. This creates an effective Hamiltonian such that particles acquire a complex phase as they hop between sites. This second method was the one used to recreate the Haldane model in reference\ \cite{jotzu14}. In this work the authors also measured the Berry phases acquired by the atom while moving through the Brillouin zone, an evidence of the topological order.

The model implemented by Jotzu et al. is topologically equivalent to the Hamiltonian that we have studied\ (\ref{eq:Haldane}) in absence of interactions: both are related by a smooth, unitary deformation of the pseudospin field that does not change the topological invariants. Our goal would be then to add interactions to the experiment, so as to observe the survival of topological invariants and the phase transitions that have been evidenced in previous sections. In\ \cite{jotzu14} the authors used\ $^{40}K$ atoms in two internal hyperfine states, \ $\ket{F, m_F} = \ket{9/2, -9/2}$ and\ $\ket{9/2, -7/2}$, but their interactions were suppressed using Feshbach resonances so that the effective model was a noninteracting version of\ (\ref{eq:Haldane}). The introduction of interactions would ``simply'' require adiabatically moving away from the Feshbach resonance. As we have shown in previous sections, the ground state in the interacting regime $|U|< |t|$, is smoothly connected to $U=0$ many-body state. We therefore expect that adiabatically increasing the interactions should not introduce significant heating.  Regarding the experimentally achievable parameters, the authors report to first neighbor hoppings to be\ $t_{FN}/h = \left[-746(81), -527(17), -527(17)\right]$Hz and the second neighbor hoppings\ $t_{SN}/h = \left[14, 14, 61\right]$Hz, where $h$ is Planck's constant. The ratio $t_{SN}/t_{FN}\sim 0.05$ is comparable to our simulations, and the experimental parameters set a limit of interaction strength of about $U \sim 3 t_{FN}\sim h\times 1.5 k$Hz.

Finally, to measure the winding of the system's ground state one has to measure first the pseudospin field,\ $\bS\left(\vk\right)$. This can be done using TOF images\ \cite{alba11,goldman13}, which reproduce  the momentum density distributions in both sublattices,\ $n_{a, b}\left(\vk\right)$, and of every spin population. The winding number\ (\ref{eq:winding}) is then straightforward to compute from the pseudospin field. 

\section{Summary and discussion}
\label{sec:discussion}

In this work we have studied the topological phase diagram of a topological insulator with spin-spin on-site interactions. To identify these topologically non-trivial phases we have estimated the winding number. Our main result is the observation that the topological phases from the noninteracting topological insulator model are extremely robust against the introduction of repulsive and attractive interactions, disappearing when interactions and hopping are comparable,\ $|U|/t \sim 3$. This prediction is supported by mean-field and MPS calculations, both of which show a remarkably good agreement in the detection of the topological phases. We have further characterized the topological phase transitions using the double occupancy\ (\ref{eq:D}) and the product of the mean pseudospin fields in the\ $z$ direction. Both order parameters showed some kind of discontinuity when the winding number changed, giving us a strong signal of a transition into a topologically trivial phase. We have also made a proposal to study these phase transitions using an existing, state-of-the-art implementation of the Haldane model\ \cite{jotzu14}, with the only requirement that interactions are reintroduced in the experiment. Finally, the remarkable agreement between our mean-field ansatz and the MPS simulations hint at the capacity of the former to capture long-range correlations and entanglement. Further work to improve the mean-field ansatz, and study time-dependent variations are in order.

\section*{Acknowledgments}

This work has been supported by Spanish MINECO Projects FIS2015-70856-P and FIS2015-63770-P (both co financed by FEDER funds) and CAM PRICYT Project QUITEMAD+ S2013/ICE-2801. The authors acknowledge the computer resources and technical assistance provided by the Centro de Supercomputaci\'on y Visualizaci\'on de Madrid (CeSViMa).

\section*{References}

\bibliography{bibi}

\end{document}